\documentclass{article}
\usepackage{natbib,epsfig,emulateapj}

\def\deg{\ifmmode^\circ\else$^\circ$\fi}

\def\h0{H$_0$}
\def\q0{q$_0$}

\def\arcs{\ifmmode {''}\else $''$\fi}
\def\arcm{\ifmmode {'}\else $'$\fi}
\def\parcs{\sa=.07em \sb=.03em
     \ifmmode $\rlap{.}$^{\scriptscriptstyle\prime\kern -\sb\prime}$\kern -\sa$
     \else \rlap{.}$^{\scriptscriptstyle\prime\kern -\sb\prime}$\kern -\sa\fi}
\def\parcm{\sa=.08em \sb=.03em
     \ifmmode $\rlap{.}\kern\sa$^{\scriptscriptstyle\prime}$\kern-\sb$
     \else \rlap{.}\kern\sa$^{\scriptscriptstyle\prime}$\kern-\sb\fi}

\def\spose#1{\hbox to 0pt{#1\hss}}
\def\simlt{\mathrel{\spose{\lower 3pt\hbox{$\mathchar''218$}}
     \raise 2.0pt\hbox{$\mathchar''13C$}}}
\def\simgt{\mathrel{\spose{\lower 3pt\hbox{$\mathchar''218$}}
     \raise 2.0pt\hbox{$\mathchar''13E$}}}
\def\lsim{\rlap{$<$}{\lower 1.0ex\hbox{$\sim$}}}
\def\gsim{\rlap{$>$}{\lower 1.0ex\hbox{$\sim$}}}

\slugcomment{To appear in ApJL; submitted Oct. 27, 1999}

\begin{document}


\title{K-Band Spectroscopy of an Obscured Massive Stellar Cluster in the
Antennae Galaxies (NGC 4038/4039) with NIRSPEC\altaffilmark{1}}

\author{
Andrea M. Gilbert\altaffilmark{2,8},
James R. Graham\altaffilmark{2}, 
Ian S. McLean\altaffilmark{3}, 
E. E. Becklin\altaffilmark{3}, 
Donald F. Figer\altaffilmark{4},
James E. Larkin\altaffilmark{3}, 
N. A. Levenson\altaffilmark{5}, 
Harry I. Teplitz\altaffilmark{6,7}, 
Mavourneen K. Wilcox\altaffilmark{3} 
}  
\altaffiltext{1}{Data presented herein were obtained
at the W.M. Keck Observatory, which is operated as a scientific
partnership among the California Institute of Technology, the
University of California and the National Aeronautics and Space
Administration.  The Observatory was made possible by the generous
financial support of the W.M. Keck Foundation.}
\altaffiltext{2}{
Department of Astronomy,  
University of California,
601 Campbell Hall,
                 Berkeley, CA, 94720-3411 }
\altaffiltext{3}{Department of Physics and Astronomy, 
University of California, 
                 Los Angeles, CA, 90095-1562 }
\altaffiltext{4}{Space Telescope Science Institute, 
                  3700 San Martin Dr., Baltimore, MD 21218 }
\altaffiltext{5}{
Department of Physics and Astronomy,  
Johns Hopkins University,
                 Baltimore, MD  21218}
\altaffiltext{6}{Laboratory for Astronomy and Solar Physics, Code 681, Goddard
Space Flight Center, Greenbelt MD 20771}
\altaffiltext{7}{NOAO Research Associate}
\altaffiltext{8}{agilbert@astro.berkeley.edu}


\begin{abstract}  
We present infrared spectroscopy of the Antennae Galaxies
(NGC~4038/4039) with NIRSPEC at the W. M. Keck Observatory.  We imaged
the star clusters in the vicinity of the southern nucleus (NGC~4039)
in $0\arcs.39$ seeing in K-band using NIRSPEC's slit-viewing
camera. The brightest star cluster revealed in the near-IR (M$_{\rm
K}(0) \simeq -17.9$) is insignificant optically, but coincident with
the highest surface brightness peak in the mid-IR ($12-18 \mu$m) ISO
image presented by \citet{mirabel98}.  We obtained high
signal-to-noise 2.03$-$2.45 $\mu$m spectra of the nucleus and the
obscured star cluster at R $\sim 1900$.

The cluster is very young (age $\sim 4$ Myr), massive (M $\sim 16
\times 10^6$ M$_{\odot}$), and compact (density $\sim 115$ M$_{\odot}$
pc$^{-3}$ within a 32 pc half-light radius), assuming a Salpeter IMF
(0.1$-$100 M$_{\odot}$).  Its hot stars have a radiation field
characterized by T$_{\rm eff}\sim 39,000$ K, and they ionize a compact
\ion{H}{2} region with n$_{\rm e}\sim 10^4$ cm$^{-3}$.  The stars are
deeply embedded in gas and dust (A$_{\rm V} \sim 9-10$ mag), and their
strong FUV field powers a clumpy photodissociation region with
densities n$_{\rm H}\ga 10^5$ cm$^{-3}$ on scales of $\sim 200$ pc,
radiating L$_{{\rm H}_2 1-0~{\rm S}(1)}= 9600$ L$_{\odot}$.

\end{abstract}


\section{Introduction}

The Antennae (NGC~4038/4039) are a pair of disk galaxies in an early
stage of merging which contain numerous massive super star clusters
(SSCs) along their spiral arms and around their interaction region
\citep{whitmore95,whitmore99}.  The molecular gas distribution peaks
at both nuclei and in the overlap region \citep{stanford90}, but the
gas is not yet undergoing a global starburst typical of more advanced
mergers \citep{nikola98}.  Star formation in starbursts appears to
occur preferentially in SSCs.  We chose to observe the Antennae
because their proximity permits an unusually detailed view of the
first generation of merger-induced SSCs and their influence on the 
surrounding interstellar medium.

The Infrared Space Observatory (ISO) $12-18~\mu$m image showed that
the hot dust distribution is similar to that of the gas, but peaks at
an otherwise inconspicuous point on the southern edge of the overlap
region \citep{mirabel98}.  This powerful starburst knot is also a
flat-spectrum radio continuum source \citep{hummel86} and may be
associated with an X-ray source \citep{fabbiano97}.  We imaged the
region around this knot, and discovered a bright compact star cluster
coincident with the mid-IR peak.  We obtained moderate-resolution (R
$\sim 1900$) K-band spectra of both the obscured cluster and the
NGC~4039 nucleus.

\section{Observations \& Data Reduction}

NIRSPEC is a new facility infrared ($0.95 - 5.6 \mu$m) spectrometer
for the Keck-II telescope, commissioned during April through July,
1999 \citep{mclean98}.  It has a cross-dispersed cryogenic echelle
with R $\sim 25,000$, and a low resolution mode with R $\sim 2000$.
The spectrometer detector is a 1024 $\times$ 1024 InSb ALADDIN focal
plane array, and the IR slit-viewing camera detector is a 256 $\times$
256 HgCdTe PICNIC array.

We observed the Antennae with NIRSPEC during the June 1999
commissioning run.  Slit-viewing camera (SCAM) images at 2 $\mu$m
reveal that the mid-IR ISO peak is a bright (K $= 14.6$) compact star
cluster located 20\arcs.4 east and 4\arcs.7 north of the K-band
nucleus.  This cluster is associated with a faint (V $= 23.5$) red
(V$-$I $= 2.9$) source (\# 80 in \citealt{whitmore95}) visible with
Space Telescope (Whitmore \& Zhang, private communication).  We
obtained low resolution (R $\simeq 1900$) $\lambda \simeq 2.03 -
2.45~\mu$m spectra through a $0\arcs.57 \times 42\arcs$ slit at
PA=77$^{\deg}$ located on the obscured star cluster and the nucleus of
NGC~4039.  The total integration time on source was 2100 s.

We dark-subtracted, mean-sky-subtracted, flat-fielded, and corrected
two-dimensional spectra for bad pixels and cosmic rays before
rectifying the curved order onto a grid in which wavelength and slit
position are perpendicular.  We then corrected for residual sky
emission and divided by a B1.5 standard star spectrum to correct for
atmospheric absorption.  The object spectra were extracted using a
Gaussian weighting function matched to their strong continuua
collapsed in wavelength (intrinsic FWHM = 0\arcs.84 for cluster,
0\arcs.99 for nucleus)\footnote{These widths are greater than those
measured from the SCAM images, $\sim 0.\arcs69$ and $\sim 0\arcs.83$
(intrinsic), due to the extended line contribution and rectification
errors of order $\la$ 1 pixel at the chip edges.}, and then an
aperture correction was applied to recover the full flux in the
continuua.  Thus we neglected more-extended H$_2$ emission, which has
maximum FWHM $\sim 1\arcs.7$ in the cluster and $\sim 1\arcs.2$ in the
nucleus.  We obtained a flux scale by requiring the 2.2 $\mu$m star
flux to equal that corresponding to its K magnitude.  Reduced spectra
are shown in Figures 1 and 2.

{\plotfiddle{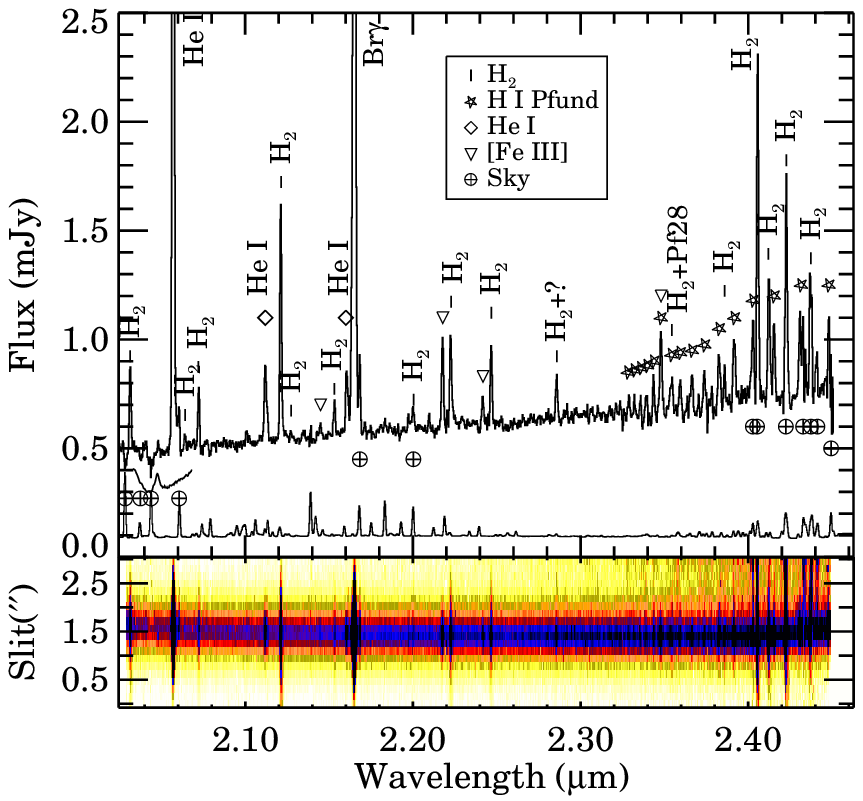}{3.2in}{0}{100}{100}{0}{10}}
{\footnotesize Figure 1. - 
NIRSPEC spectrum of the obscured star
cluster shows nebular and fluorescent H$_2$ emission with a continuum
rising toward the red.  Scaled sky counts are plotted at 0.1 mJy.
$\omega$-shaped curve represents an atmospheric CO$_2$ band at 2.05
$\mu$m. 
}
\vspace{-0.08in}

\section{Massive Star Cluster}

The cluster spectrum is characterized by strong emission
lines\footnote{A table of measured line fluxes is available
electronically from 
{http://astro.berkeley.edu/$\sim$agilbert/antennae}.} and
a continuum (detected with SNR $\simeq 15$) dominated by the light of
hot, blue stars and dust.  The nebular emission lines are slightly
more extended than the continuum, and the H$_2$ emission is even more
extended.  This suggests a picture in which hot stars and dust are
embedded in a giant compact H~{\sc ii} region surrounded by clumpy
(see \S 3.2) clouds of obscuring gas and dust whose surfaces are
ionized and photodissociated by FUV photons escaping from the star
cluster.

For a distance to the Antennae of 19 Mpc (H$_0$=75 km s$^{-1}$
Mpc$^{-1}$, 1\arcs = 93 pc) \citep{whitmore99}, we find that the
cluster has M$_{\rm K} = -16.8$.  We estimate the screen extinction to
the cluster by assuming a range of (V$-$K)$_0 \simeq 0-1$ as expected
from Starburst99 models \citep{leitherer99}, and that A$_{\rm K} =
0.11$ A$_{\rm V}$ \citep{rieke85}.  We find A$_{\rm V} = 9-10$ mag,
which implies M$_{\rm K}(0) = -17.9$, adopting A$_{\rm K}=1.1$ (which
is confirmed by our analysis of the \ion{H}{2} recombination lines in
\S 3.1).  We can use the intrinsic brightness along with the Lyman
continuum flux inferred from the de-reddened Br$\gamma$ flux ($3.1
\times 10^{-14}$ erg s$^{-1}$ cm$^{-2}$), Q(H$^+)_0 = 1.0 \times
10^{53}$ photons s$^{-1}$, to constrain the cluster mass and age.
Using instantaneous Starburst99 models we find a total mass of $\sim 7
\times 10^6$ M$_{\odot}$ (with $\sim 2600$ O stars) for a Salpeter IMF
extending from 1 to 100 M$_\odot$, and an age of $\sim 4$ Myr.  This
age is consistent with the lack of photospheric CO and metal
absorption lines from red supergiants and other cool giants, which
would begin to contribute significantly to the 2 $\mu$m light at an
age of $\sim 7$ Myr \citep{leitherer99}.  The cluster's density is
then about 115 M$_{\odot}$ pc$^{-3}$ for stars of 0.1$-$100
M$_{\odot}$ within a half-light radius of $32$ pc.  This density is 30
times less than that of the LMC SSC, R136 (within a radius of 1.7 pc,
assuming a Salpeter proportion of low-mass stars) \citep{hunter95}.
Thus the Antennae cluster may be a complex of clusters rather than one
massive cluster.

{\plotfiddle{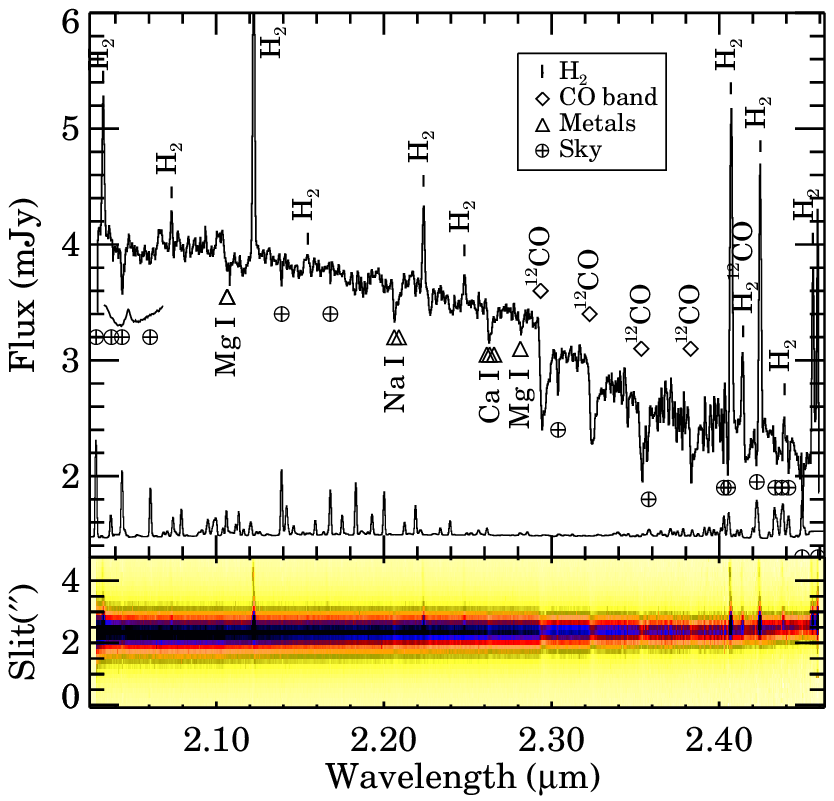}{3.2in}{0}{100}{100}{0}{10}}
{\footnotesize Figure 2. - 
NIRSPEC spectrum of NGC~4039 nucleus shows
extended collisionally excited H$_2$ emission and a strong stellar
continuum marked by photospheric absorption.  No Br$\gamma$ is present.
Scaled sky counts are shown at 1.5 mJy.
}
\vspace{-0.1in}

\subsection{Nebular Emission}

The cluster spectrum features a variety of nebular lines that reveal
information about the conditions in the ionized gas around the
cluster, which in turn allows us to constrain the effective
temperature of the ionizing stars.  

We detected \ion{H}{1} Pfund series lines from Pf~19 to Pf~38, and
display their fluxes relative to that of Br$\gamma$ in Figure 3.
The filled symbols give fluxes for the blends Pf
28+H$_2$ 2$-$1 S(0) and Pf 29+[\ion{Fe}{3}].  They fall well above
the other points, which follow closely the theoretical expectation for
intensities relative to Br$\gamma$ (solid curve) with no reddening
applied, for a gas with n$_{\rm e} = 10^{4}$ cm$^{-3}$ and T$_{\rm
e}=10^{4}$ K \citep{hummer87}.  Excluding the two known blends, the
best-fit foreground screen extinction is A$_{\rm K}=1.1 \pm 0.3 $ mag
(dashed curve), assuming the extinction law of \citet{landini84} and
evaluated at 2.2 $\mu$m.  We consider this an upper limit on A$_{\rm
K}$ because a close look at the spectrum shows that the points above the
dashed line in Figure 3
for Pf 22$-$24 at 2.404, 2.393, and
2.383 $\mu$m may also be blended or contaminated by sky emission,
implying a lower A$_{\rm K}$ and a much better fit to the theory.
Hence the majority of the extinction to the cluster is bypassed by
observing it in K band.  

\vspace{-0.06in}
{\plotfiddle{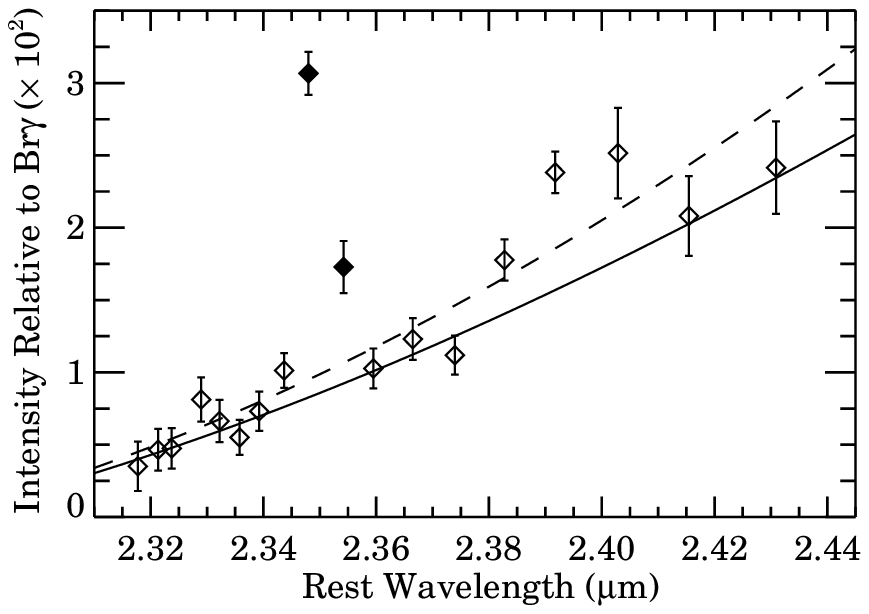}{2.5in}{0}{100}{100}{0}{5}}
{\footnotesize Figure 3. - 
Pfund line fluxes relative to Br$\gamma$
flux ($1.05 \times 10^{-14}$ erg s$^{-1}$cm$^{-2}$).  Solid curve is
unextincted theoretical curve for n$_{\rm e} = 10^{4}$ cm$^{-3}$,
T$_{\rm e}=10^{4}$ K (Hummer \& Storey 1987).  Filled symbols
represent lines that are known blends, and the dashed curve shows
theoretical fluxes with the best-fit extinction A$_{\rm K}=1.1$
mag.
}
\vspace{0.04in}

The lack of a strong Pfund discontinuity at 2.28 $\mu$m indicates that
nebular free-free and bound-free continuum is diluted by starlight
and dust emission (signaled by the rising continuum toward longer
$\lambda$) in the cluster.

The ratios of [\ion{Fe}{3}] lines are nebular density diagnostics;
Table 1
presents observed ratios and theoretical
predictions of \citet{keenan92} for emission from a collisionally
excited 10$^4$ K gas, as tabulated by \citet{luhman98}.  The ratios of
[\ion{Fe}{3}] 2.146 $\mu$m and [\ion{Fe}{3}] 2.243 $\mu$m to
[\ion{Fe}{3}] 2.218 $\mu$m are consistent with n$_{\rm e}= 10^{3.5} -
10^4$ cm$^{-3}$.  The ratio [\ion{Fe}{3}] 2.348 $\mu$m/[\ion{Fe}{3}]
2.218 $\mu$m is 20\% higher than its theoretical value, which is
roughly constant over all of parameter space \citep{keenan92}, but
[\ion{Fe}{3}] 2.348 $\mu$m is blended with Pf 29 and subject to
measurement errors that are larger than the difference in extinctions
in question (see Figure 3).
Even the minimum value we infer for this ratio, with $A_{\rm K}=0$, is
significantly greater than the model prediction.  High values of
[\ion{Fe}{3}] 2.348 $\mu$m/[\ion{Fe}{3}] 2.218 $\mu$m were also
found by \citet{luhman98} in Orion.  This discrepancy may be due to
blending with another unknown line, or to theoretical error; ratios
from the latest calculations have an average deviation from data of
10\% \citep{keenan92}.

\ion{He}{1} line ratios can in principal be used to infer nebular
temperature T$_{\rm e}$, and are fairly insensitive to n$_{\rm e}$.
However, of the three lines we detected, two are not suitable for such
an analysis: the \ion{He}{1} 2.1615+2.1624 $\mu$m blend falls on the
wing of strong Br~$\gamma$ so its flux has a large (50\%) measurement
error, and the strong \ion{He}{1} 2.0589 $\mu$m line is subject to
radiative transfer and density effects.

The \ion{He}{1} 2.0589 $\mu$m/Br~$\gamma$ ratio is an indicator of the
T$_{\rm eff}$ of hot stars in \ion{H}{2} regions \citep{doyon92},
although it is sensitive to nebular conditions such as the relative
volumes and ionization fractions of He$^+$ and H$^+$, geometry,
density, dustiness, etc. \citep{shields93}. \citet{doherty95} studied
H and He excitation in a sample of starburst galaxies and \ion{H}{2}
regions.  For starbursts they found evidence for high-T$_{\rm eff}$,
low-n$_{\rm e}$ ($\sim 10^2$ cm$^{-2}$) ionized gas from \ion{He}{1}
2.0589 $\mu$m/Br~$\gamma$ ratios of 0.22 to 0.64.  This is consistent
with giant extended \ion{H}{2} regions expected to dominate the
emission-line spectra of typical starbursts.  The ultra-compact
\ion{H}{2} regions were characterized by higher ratios (0.8$-$0.9) and
higher densities, $\sim$ 10$^4$ cm$^{-3}$.  The cluster has a flux
ratio of 0.70, a value between the two object classes of
\citet{doherty95}.  Assuming the line emission is purely nebular, this
ratio is consistent with a high-density (10$^4$ cm$^{-3}$) model of
\citet{shields93}, and implies T$_{\rm eff}\simeq 39,000$ K for the
assumed model parameters.  This temperature is similar to that derived
by \citet{kunze96}, $\simeq 44,000$ K, from mid-IR SWS line
observations in a large aperture on the overlap region of the
Antennae.

\vspace{-0.05in}
\begin{center}
{\centering
\begin{tabular}{lccrcc}
\multicolumn{6}{c}{TABLE 1} \\
\multicolumn{6}{c}{\sc Cluster [\ion{Fe}{3}] Line Ratios\tablenotemark{a}} \\
\tableline
\tableline
\multicolumn{1}{c}{} & Rest & Observed & \multicolumn{3}{c}{Model
Ratio\tablenotemark{c}} \\
\cline{4-6}
\multicolumn{1}{c}{Transition} & $\lambda$($\mu$m)\tablenotemark{b} & Ratio & 
$10^3$ & $10^4$  & $10^5$ \\
\tableline
$^3$G$_{3}-^3$H$_4$ &  2.1457 &  0.14$\pm$0.02  & 0.10 & 0.17 & 0.34 \\
$^3$G$_{5}-^3$H$_6$ &  2.2183 &  1.00           & 1.00 & 1.00 & 1.00 \\
$^3$G$_{4}-^3$H$_4$ &  2.2427 &  0.28$\pm$0.02  & 0.26 & 0.29 & 0.38 \\
$^3$G$_{5}-^3$H$_5$ &  2.3485 &  0.80$\pm$0.03\tablenotemark{d} & 0.66 & 0.66 & 0.66 \\           
\tableline
\end{tabular}}
\end{center}
\noindent
{\footnotesize $^{\rm a}$
Ratios are dereddened fluxes relative to
[Fe~{\sc iii}] 2.2183 $\mu$m, for which the dereddened flux was
9.11$\times$10$^{-16}$ ergs s$^{-1}$ cm$^{-2}$.}\\
{\footnotesize $^{\rm b}$
\citet{sugar85}.} \\
{\footnotesize $^{\rm c}$
Models for T$_{\rm e}$=10$^4$K, values of n$_{\rm e}$ in cm$^{-3}$
\citep{keenan92}.}\\
{\footnotesize $^{\rm d}$
Flux determined by subtracting Pf 29 contribution
obtained for the best-fit Landini extinction curve 
with A$_{\rm K}=1.1$ mag.}\\
\vspace{-0.05in}

The cluster has properties more like those of a compact \ion{H}{2}
region than a diffuse one.  It appears to be a young, hot,
high-density \ion{H}{2} region, one of the first to form in this part
of the Antennae interaction region (see \citealt{habing79} for a
review of compact \ion{H}{2} regions).

\subsection{Molecular Emission}

The spectrum shows evidence for almost pure UV fluorescence excited by
FUV radiation from the O \& B stars; the strong, vibrationally excited
1$-$0, 2$-$1 \& 3$-$2 H$_2$ emission has T$_{\rm vib} \ga 6000$~K and
T$_{\rm rot}\simeq 970$, 1600, and 1800 K, respectively, and weak
higher-v (6$-$4, 8$-$6, 9$-$7) transitions are present as well.  The
H$_2$ lines are extended over $\simeq 200$~pc, about twice the extent
of the continuum and nebular line emission, so a significant fraction
of the FUV (912$-$1108 \AA) light escapes from the cluster to heat and
photodissociate the local molecular ISM.

We obtained the photodissociation region (PDR) models of
\citet{draine96} and compared them with our data by calculating
reduced $\chi_\nu^2$.  Models with high densities (n$_{\rm H} = 10^5$
cm$^{-3}$), moderately warm temperatures (T $= 500$ to 1500 K at the
cloud surface), and high FUV fields (G$_0 = 10^3 - 10^5$ times the mean
interstellar field) can reasonably fit the data.
Figure 4
shows $\chi_\nu^2$ contours for all models projected onto the n$_{\rm
H} -$G$_0$ plane.  The best-fit Draine \& Bertoldi model is n2023b,
which has n$_{\rm H}$ = 10$^5$ cm$^{-3}$, T = 900 K, and G$_0=5000$.
We fit 22 H$_2$ lines, excluding 3$-$2 S(2) 2.287 $\mu$m because it
appears to be blended with a strong unidentified nebular line at 2.286
$\mu$m found in higher-resolution spectra of planetary nebulae
\citep{smith81}.  The weak high-v transitions are all under-predicted
by this model, and appear to come from lower-density gas (n$_{\rm H}
\la 10^3-10^4$ cm$^{-3}$) exposed to a weaker FUV field (G$_0 \la
10^2-10^3$).

The ortho/para ratio of excited H$_2$ determined from the relative
column densities in v=1, J=3 and J=2 inferred from 1$-$0 S(1) and S(0)
lines is 1.62$\pm$0.07.  This is consistent with the ground state v=0,
J=1 and J=0 H$_2$ being in LTE with ortho/para ratio of 3 if the FUV
absorption lines populating the non-LTE excited states are optically
thick \citep{sternberg99}.  Indeed, the best-fit PDR models have
temperatures that are comparable with T$_{\rm rot}$ in the lowest
excited states, as well as with the warm gas kinetic temperature in
the Galactic PDR M16, T = 930$\pm$50 K, measured by
\citet{levenson99}.

If the extent of the H$_2$ emission indicates that the mean-free path
of a FUV photon is $\sim 200$ pc, then $\langle$n$_{\rm H}\rangle$ = 3
cm$^{-3}$ for a Galactic gas-to-dust ratio, while in the PDR(s)
n$_{\rm H}$ = 10$^4-10^6$ cm$^{-3}$.  This implies that the molecular
gas is extremely clumpy, which is consistent with the range of
densities inferred from the detection of anomalously strong v = 8$-$6
H$_2$ emission.

\section{NGC 4039 Nucleus}

The spectrum of the nucleus of NGC~4039 is marked by strong stellar
continuum and bright, extended H$_2$ emission.  Strong photospheric
\ion{Mg}{1}, \ion{Na}{1}, \ion{Ca}{1} absorption and CO $\Delta v=2$
bands indicate that the continuum is dominated by old giants.  The CO
band head is stronger than that of a M2III, suggesting some
contribution from red supergiants.  The absence of Br$\gamma$ emission
implies that star formation is currently extinct in the nucleus.
Spatially extended, collisionally excited H$_2$ emission in the
nucleus may be excited by SNR shocks from the last generation of
nuclear star formation, or by merger-induced cloud collisions.  We
defer detailed analysis of the nuclear spectrum to a later paper.

\section{Conclusions}

The highest surface brightness mid-IR peak in the ISO map of the
Antennae Galaxies is a massive ($\sim 16\times$10$^6$ M$_{\odot}$),
obscured (A$_{\rm V} \sim 9-10$), young (age $\sim 4$ Myr) star
cluster with half-light radius $\sim$ 32 pc, whose strong FUV flux
excites the surrounding molecular ISM on scales of up to 200 pc.  The
cluster spectrum is dominated by extended fluorescently excited
H$_{2}$ emission from clumpy PDRs and nebular emission from compact
\ion{H}{2} regions.  In contrast, the nearby nucleus of NGC~4039 has a
strong stellar spectrum dominated by cool stars, where the only
emission lines are due to shock-excited H$_2$.  These observations
confirm the potential of near-infrared spectroscopy for exploration
and discovery with the new generation of large ground-based
telescopes.  Our ongoing program of NIRSPEC observations promises
to reveal a wealth of information on the nature of star formation in
star clusters.

\vspace{-0.05in}
{\plotfiddle{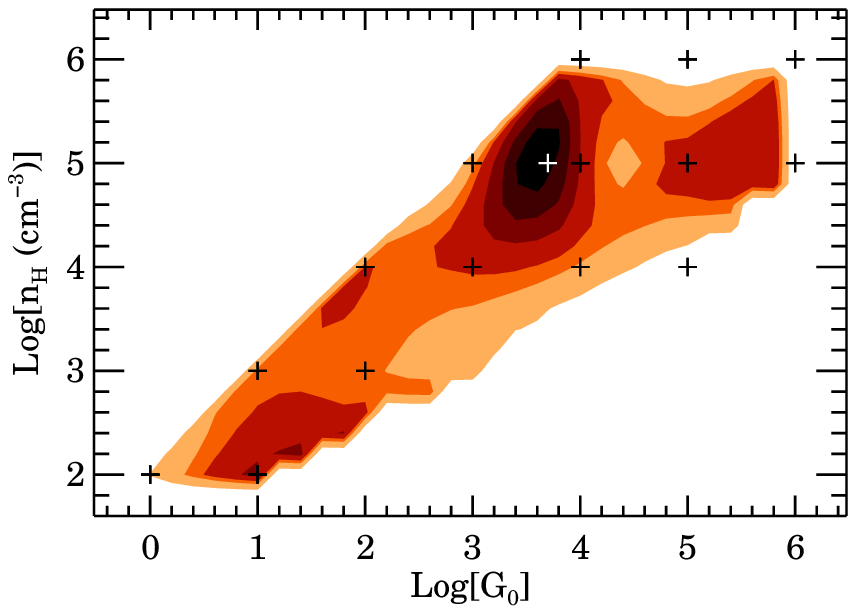}{2.55in}{0}{100}{100}{0}{10}} 
{\footnotesize Figure 4. - 
Comparison of H$_2$ line strengths with PDR models.
Contours of $\chi^2_\nu$ for 22 lines projected onto n$_{\rm
H} -$G$_0$ plane peak at n$_{\rm H} \sim 10^5$ cm$^{-3}$ and G$_0 \sim
5000$. Model points (+) are for T$_0$ = 300 $-$ 2000
K. White + marks best-fit PDR model of Draine \& Bertoldi, with T$_0$
= 900 K and $\chi_\nu^2=9.3$.  Contours are 50, 25, 20, 15, 12, 10.
}

\acknowledgements 

We acknowledge the hard work of past and present members of the UCLA
NIRSPEC team: M. Angliongto, O. Bendiksen, G. Brims, L. Buchholz,
J. Canfield, K.  Chin, J. Hare, F. Lacayanga, S. Larson, T. Liu, N.
Magnone, G. Skulason, M. Spencer, J. Weiss and W.  Wong.  We thank
Keck Director Chaffee and all the CARA staff involved in the
commissioning and integration of NIRSPEC, particularly instrument
specialist T. Bida. We especially thank Observing Assistants
J. Aycock, G. Puniwai, C. Sorenson, R. Quick and W. Wack for their
support.  We also thank A. Sternberg for valuable discussions.  We are
grateful to R. Benjamin for providing us with He~{\sc i} emissivity
data.  AMG acknowledges support from a NASA GSRP grant.

\end{document}